\documentclass[twocolumn,prl,floatfix,citeautoscript,nofootinbib,superscriptaddress]{revtex4}
\usepackage{eurosym}
\usepackage{amsfonts}
\usepackage{mathrsfs}
\usepackage{xspace}
\usepackage{subfigure}
\usepackage{longtable}
\usepackage{xspace}
\usepackage{multirow}
\usepackage{tabularx}
\usepackage{amsmath}
\usepackage{amssymb}
\usepackage{graphicx}
\usepackage{dcolumn}
\usepackage{bm}
\usepackage[colorlinks,urlcolor=blue,citecolor=blue,linkcolor=blue]{hyperref}

\begin{document}

\title{Experimental realization of a superfluid stripe phase in a
spin-orbit-coupled Bose-Einstein condensate enabled by momentum-space hopping%
}
\author{Thomas M. Bersano}
\affiliation{Department of Physics and Astronomy, Washington State University, Pullman,
Washington 99164-2814}
\author{Junpeng Hou}
\affiliation{Department of Physics, The University of Texas at Dallas, Richardson, Texas
75080-3021, USA}
\author{Sean Mossman}
\affiliation{Department of Physics and Astronomy, Washington State University, Pullman,
Washington 99164-2814}
\author{Vandna Gokhroo}
\affiliation{Department of Physics and Astronomy, Washington State University, Pullman,
Washington 99164-2814}
\author{Xi-Wang Luo}
\affiliation{Department of Physics, The University of Texas at Dallas, Richardson, Texas
75080-3021, USA}
\author{Kuei Sun}
\affiliation{Department of Physics, The University of Texas at Dallas, Richardson, Texas
75080-3021, USA}
\author{Chuanwei Zhang}
\thanks{Email address: chuanwei.zhang@utdallas}
\affiliation{Department of Physics, The University of Texas at Dallas, Richardson, Texas
75080-3021, USA}
\author{Peter Engels}
\thanks{Email address: engels@wsu.edu}
\affiliation{Department of Physics and Astronomy, Washington State University, Pullman,
Washington 99164-2814}

\begin{abstract}
In the past few decades, the search for supersolid-like phases has attracted
great attention in condensed matter and ultracold atom communities. Here we
experimentally demonstrate a route for realizing a superfluid stripe-phase
in a spin-orbit coupled Bose-Einstein condensate by employing a weak optical
lattice to induce momentum-space hopping between two spin-orbit band minima.
We characterize the striped ground state as a function of lattice coupling
strength and spin-orbit detuning and find good agreement with mean-field
simulations. We observe coherent Rabi oscillations in momentum space
between two band minima and demonstrate a long lifetime of the ground state.
Our work offers an exciting new and stable experimental platform for
exploring superfluid stripe-phases and their exotic excitations, which may
shed light on the properties of supersolid-like states.
\end{abstract}

\maketitle

{\color{blue}\emph{Introduction}}. Supersolids are an exotic phase of matter
which simultaneously possess the crystalline properties of a solid and the
unique flow properties of a superfluid \cite{Boninsegni2012}. Such
simultaneous breaking of continuous translational symmetry and U(1) gauge
symmetry was first predicted for solid helium \cite{Thouless1969,Andreev1971}%
, but convincing evidence of a supersolid state in this system has remained elusive \cite%
{Kim2012}. In recent years, the experimental realization of spin-orbit
coupling (SOC) in ultracold atomic gases \cite%
{Lin2011,Zhang2012b,Qu2013a,Olson2014,
Hamner2014,Wang2012,Cheuk2012,Williams2013,Lev,Jo,Huang2016,
Meng2016,Pan2016} has opened a new pathway for demonstrating long-sought
supersolid-like states \cite{Stanescu2008, Wu2011, wang2010spin, ho2011bose,
li2012quantum, zhang2012mean, hu2012spin, ozawa2012stability,
sun2016interacting, yu2016phase, martone2016tricriticalities,Luo2017,Hou2018}%
.

The lowest energy band in the SOC dispersion is characterized by two local
minima at distinct momenta \cite{Lin2011}. For a narrow range of system
parameters, mean-field interactions within a Bose-Einstein condensate (BEC)
favor a ground state which is composed of a coherent superposition of two
plane-wave states at the dispersion minima \cite{li2012quantum}. This
superposition leads to density modulations in real space, or stripes,
therefore breaking translational symmetry while maintaining the superfluid
phase correlation of a BEC. Such a stripe-phase was initially proposed for SOC
BECs where the pseudospins are defined by two atomic hyperfine states \cite%
{Lin2011}. While great experimental progress has been made in exploring the
rich physics of such SOC systems, a ground state superfluid stripe-phase has
not been observed in this context. The necessary parameter space is
prohibitively sensitive to magnetic field fluctuations and the resulting
density modulation is weak. However, recent works have attempted to sidestep
these difficulties in creative ways, leading to experimental observations of
some signatures of superfluid stripe phases in different systems \cite%
{Li2016,Li2017,Leonard2017,Chomaz2018}.

Despite these significant advances, the quest for a robust and long-lived
platform for the experimental investigations of stripe-phase properties
remains. In this Letter, we show that the superposition of two local band
minima to form a supersolid-like ground state can be robustly achieved by
means other than atomic interactions. Specifically, we engineer
momentum-space hopping between two local band minima in a SOC BEC driven by
a weak optical lattice \cite{Martone2016,Liu2016,Hou20182}. The
coupling between different momenta through static or moving optical lattices
has been widely used in ultracold atomic gases for engineering versatile types of
physics \cite{Andersen2006,Engels2016,Meier2016,An2018}, but momentum states
involved in such coupling schemes are usually not local band minima and thus
have short lifetime. Here such a coupling is applied between two local band
minima, hence denoted as hopping, in analogy to the hopping between real-space
double wells.

When the decrease of the energy due to momentum-state hopping exceeds the
increase of the interaction energy due to density modulation, a stripe-phase
is more favorable than a plane-wave state at a single dispersion minimum \cite{Supp}.
In a scheme where SOC is induced by Raman coupling two atomic hyperfine states,
this condition can be easily satisfied even for large Raman coupling
strengths and detunings, yielding a large parameter region with strong
density modulation and spin mixture. While the breaking of continuous
translational symmetry is triggered by a weak optical lattice, rather than
spontaneously broken, the resulting stripe-phase shares similar physics with
an authentic supersolid state \cite{Martone2017}. In this sense, the addition of momentum
hopping to SOC offers a robustly tunable and long-lived experimental
platform for exploring exotic superfluid stripe-phase excitations (e.g.
breathing modes, roton excitations and collective dynamics \cite{Chen2012}),
which may shed light on the properties of supersolid-like states.

In our experiments, we first verify that the spin and momentum composition
of our experimentally realized state is consistent with the expected ground
state stripe-phase. We then probe the coherent momentum hopping between two
SOC dispersion minima by jumping on a weak optical lattice to induce Rabi
oscillations. Finally, we investigate the coherence and lifetime of the
ground state stripe-phase by quenching the SOC detuning at different times
and observing the population dynamics. These experimental observations of
ground states and dynamics are in good agreement with the numerical
simulations of the Gross-Pitaevskii equation (GPE), therefore providing
significant evidence for the successful production of a stripe-phase ground
state.

\begin{figure}[t]
\includegraphics[width=0.48\textwidth]{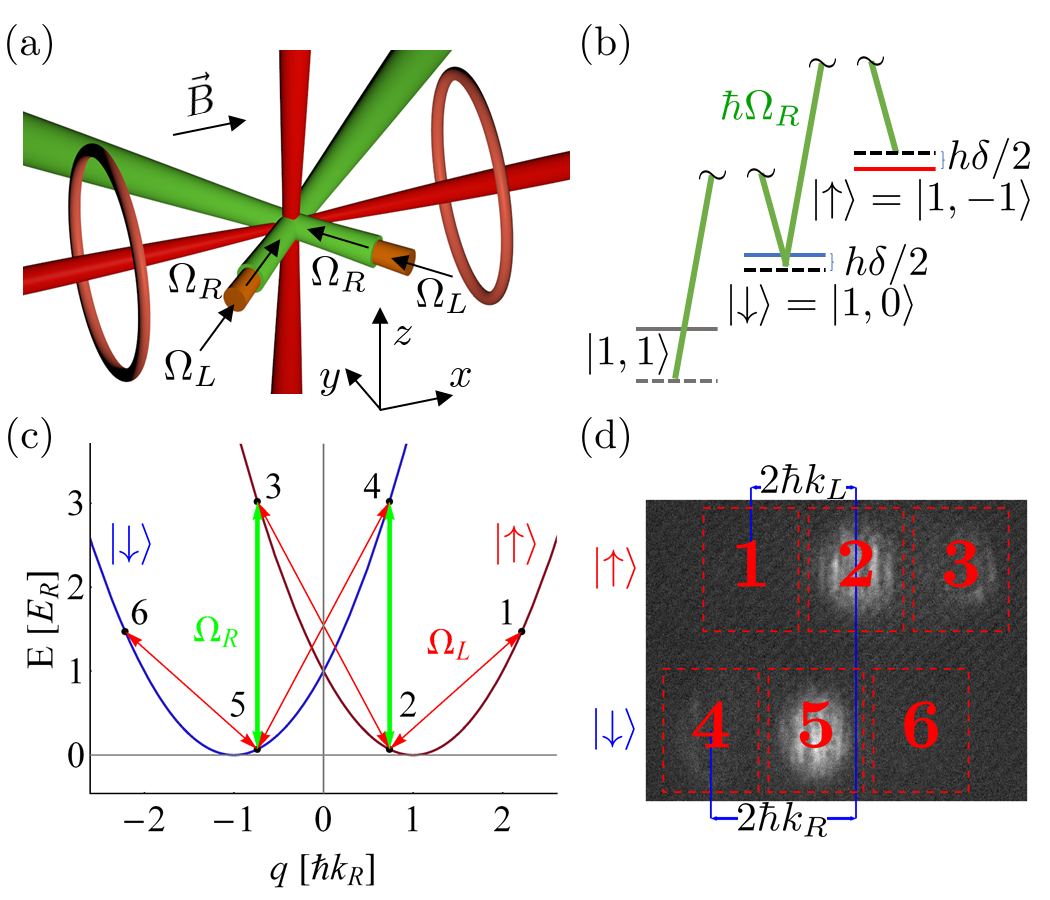}
\caption{(a) Schematic of the experimental configuration. A BEC is held in a
crossed dipole trap (red) in the presence of a 10 Gauss bias field. Two
colinear Raman beams (green) and lattice beams (red) intersect at the BEC
position. (b) The Raman coupling scheme. The Raman beams couple the $\lvert
1,-1\rangle $ and $\lvert 1,0\rangle $ states of the $F=1$ manifold with
detuning $\protect\delta $. (c) The Raman coupling (green arrows) and the
lattice coupling (red arrows) of the relevant momentum states in a rotated
spin basis at zero detuning. (d) An example of an experimental measurement
using absorption imaging after time-of-flight through a Stern-Gerlach field.
The bare states are enumerated to correspond with the marking on the
bare-state coupling scheme in (c). Spin-up atoms appear above
spin-down atoms due to the Stern-Gerlach field.
The occupied momentum states separate horizontally during the time of flight.}
\label{fig:experiment}
\end{figure}

{\color{blue}\emph{Spin-orbit coupling with lattice-assisted hopping}}. We
consider the experimental geometry shown in Fig.~\ref{fig:experiment}(a).
A BEC is prepared in a crossed dipole trap.
Two Raman beams are incident on the BEC at 45$^{\circ }$ angles relative to
the $x$-axis, inducing SOC along the $x$-direction. These Raman beams couple
two atomic spin states $\lvert 1,-1\rangle $ and $\lvert 1,0\rangle $ within
the $F=1$ hyperfine manifold, which we will refer to as pseudospin $\lvert
\uparrow \rangle $ and ${\lvert \downarrow \rangle }$ respectively, as shown
in Fig.~\ref{fig:experiment}(b). This two-photon Raman transition is detuned
by $\delta $, while the third spin state ($|1,1\rangle $) is far
off-resonance due to the quadratic Zeeman shift, leaving it effectively
decoupled from the other two states.

The coupled spin states are separated in momentum space by $2\hbar k_{R}$
where the Raman recoil momentum $\hbar k_{R}$ is the wavevector of the Raman
beams projected onto the $x$-axis. In a rotated basis, the states are
described by a quasi-momentum given by $q=p\pm \hbar k_{R}$ for the two spin
states, where $p$ is the free-particle momentum. The SOC can be visualized
as vertical transitions (green) in Fig.~\ref{fig:experiment}(c).
Diagonalization of the SOC Hamiltonian results in a two-band structure
where, for suitable parameters, the lower spin-orbit band features two
minima that are located at $q_{\text{min}}=\pm \hbar k_{R}\sqrt{1-(\frac{%
\hbar \Omega _{R}}{4E_{R}})^{2}}$ in single-particle regimes \cite%
{li2012quantum}. Here $\hbar \Omega _{R}$ represents the Raman coupling
strength, the recoil energy is $E_{R}=\hbar ^{2}k_{R}^{2}/2m$, and $m$ is
the atomic mass of $^{87}$Rb.

In addition to the Raman beams, two lattice beams copropagating with the
Raman beams create an optical lattice potential $V_{L}=2\hbar \Omega
_{L}\sin ^{2}(k_{L}x)$ along the $x$-direction. This static spin-independent lattice can provide
a $2\hbar k_{L}$ momentum-kick while conserving the spin \cite{Supp}. Such an effect is illustrated
by the diagonal couplings (red arrows) marked in Fig.~\ref{fig:experiment}%
(c). The key feature of this configuration is that the wavelength of the
lattice beams and the Raman coupling strength are chosen such that the
lattice couples the two minima of the lower spin-orbit band, i.e. $\hbar
k_{L}=|q_{\text{min}}|$.

The dynamics of this system can be described by a one-dimensional
GPE under the mean-field approximation,
\begin{equation}
i\hbar \frac{\partial }{\partial t}\psi =\left( H_{\text{SOC}}\ +V_{L}+\frac{%
1}{2}m\omega _{x}^{2}x^{2}+\frac{g}{2}|\psi |^{2}\right) \psi ,
\end{equation}%
where $\omega _{x}$ is the trapping frequency along $x$ and $\psi =(\psi
_{\uparrow },~\psi _{\downarrow })^{T}$ is the two-component spinor
wavefunction normalized by the average particle number density $n=\int
dx\psi ^{\dagger }\psi $. The SOC term is
\begin{equation}
H_{\text{SOC}}=\left(
\begin{array}{cc}
\frac{1}{2m}(q_{x}-\hbar k_{R})^{2}-\frac{h\delta }{2} & \frac{1}{2}\hbar
\Omega _{R} \\
\frac{1}{2}\hbar \Omega _{R} & \frac{1}{2m}(q_{x}+\hbar k_{R})^{2}+\frac{%
h\delta }{2}%
\end{array}%
\right) .
\end{equation}%
Although it is not obvious in the bare momentum basis, the two spin-orbit
band minima are directly coupled in the SOC dressed basis by the weak
optical lattice, which we name as ``\textit{momentum-space hopping}".

\begin{figure}[t]
\includegraphics[width=0.48\textwidth]{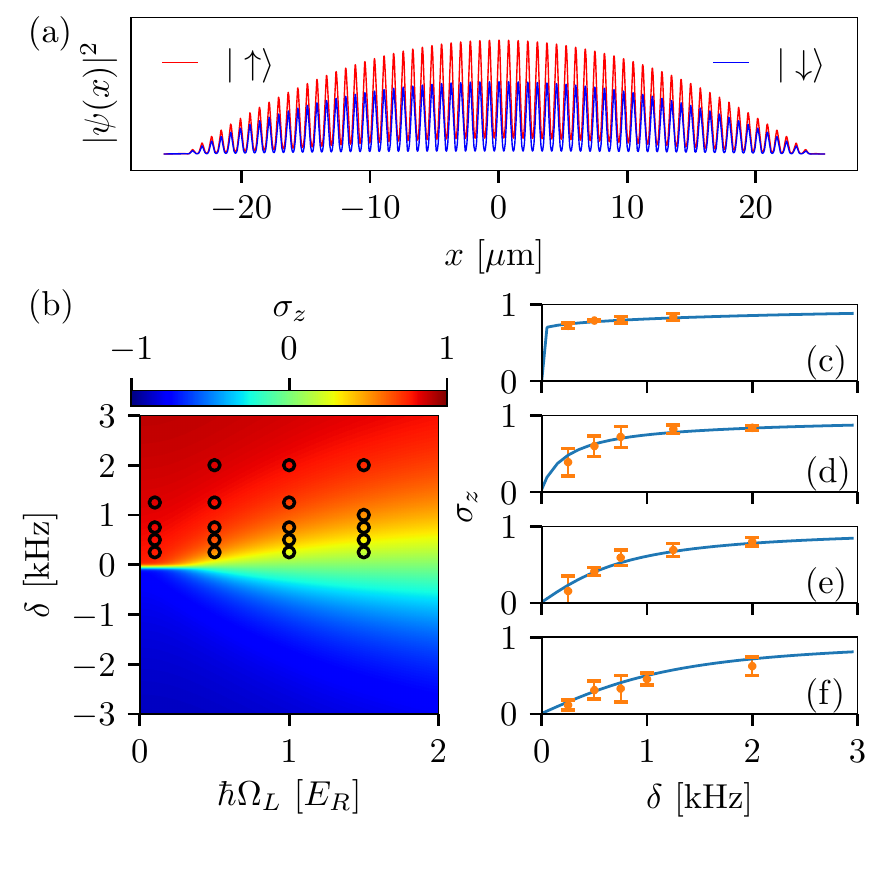}
\caption{(a) A numerical example of the stripe ground state density
modulations with $\lvert \uparrow \rangle $ shown in red and $\lvert
\downarrow \rangle $ shown in blue where $\hbar \Omega _{R}=2.7E_{R}$, $\protect%
\delta =250$ Hz, and $\hbar \Omega _{L}=1E_{R}$. (b) Numerical simulation of
the ground state spin-polarization phase diagram as a function of SOC
detuning $\protect\delta $ and lattice coupling strength $\hbar \Omega _{L}$%
. The circles indicate the locations experimentally probed in the following subfigures. (c)-(f)
Experimental data (dots) and numerical prediction (lines) of spin
polarization as a function of Raman detuning for fixed lattice strengths of $%
\hbar \Omega _{L}=$ 0.1(c), 0.5(d), 1.0(e), and 1.5 $E_{R}$(f) respectively. All
numerics and experiments are done with SOC coupling strength $\hbar \Omega
_{R}=2.7E_{R}$. Experimental data points are averages over four measurements
with error bars given by statistical error.}
\label{fig:PhaseDiagram}
\end{figure}

{\color{blue}\emph{Ground state phase diagram}}. The momentum space hopping
gives rise to strong density modulations in real-space resulting from a
coherent population of both band minima at quasimomenta $\pm \hbar k_{L}$. A
typical ground state profile from GPE simulation is presented in Fig.~\ref%
{fig:PhaseDiagram}(a). In the absence of SOC, such high density modulation would
usually be achieved in deep optical lattices with lattice depths $\hbar \Omega _{L}\gtrsim 10E_{R}$,
while here $\hbar \Omega _{L}\approx E_{R}$ and the strong modulation
mainly originates from large Raman coupling \cite{Supp}. Given the separation
of the momentum minima $2q_{\text{min}}$, the real-space distance between
the modulation peaks is $\pi /q_{\text{min}}=0.76~\mu $m, which is confirmed
in the GPE simulations (Fig.~\ref{fig:PhaseDiagram}(a)). Experimentally, we
cannot directly observe density modulations at this length scale due to the
optical resolution limit of our imaging configuration. Instead, we infer the
existence of the stripe-phase ground state by verifying other observable
parameters, such as the spin polarization of the ground state and the
frequency of Rabi oscillations.

We begin by exploring the ground state spin polarization as a function of
lattice coupling strength $\hbar \Omega _{L}$ and SOC detuning $\delta $. After
adiabatically dressing the BEC with SOC at a large detuning $\delta =5$ kHz,
the lattice beams are ramped on over 50 ms and then the Raman detuning is
adiabatically lowered to a final value. The resulting ground state is imaged
and the spin-polarization is measured. The total spin-polarization of the
system is given by $\sigma _{z}=\sum\limits_{i}{\ s_{i}N_{i}}%
/\sum\limits_{i}\left\vert s_{i}\right\vert N_{i}$, where $N_{i}$ is the
number of atoms in state $i$ with pseudo-spin $s_{i}$, which we take to be $%
1/2$ and $-1/2$ for atoms in hyperfine levels $|1,-1\rangle $ and $%
|1,0\rangle $ respectively. These populations $N_{i}$ are determined through
Stern-Gerlach imaging after expansion such that the momentum states separate
horizontally, while the spin states separate vertically, as shown in Fig.~%
\ref{fig:experiment}(d).

The experimental data are presented along with the results from numerical
simulations in Fig.~\ref{fig:PhaseDiagram}(b-f). The numerically determined
ground state phase diagram as a function of SOC detuning and lattice
strength, shown in Fig.~\ref{fig:PhaseDiagram}(b), reveals that for small
ratios of $|\hbar \Omega _{L}/\delta |$, nearly all atoms occupy one
spin state. As the ratio increases, the two band minima become more evenly
populated and the spin-polarization approaches zero. Intuitively, we see a finite
SOC detuning energetically favors one spin-state while the lattice coupling
mixes the spin-momentum states. A closer observation reveals that the spin
polarization changes smoothly (Fig.~\ref{fig:PhaseDiagram}(d-f)) with
respect to detuning where the lattice coupling is strong, but shows an
abrupt change (Fig.~\ref{fig:PhaseDiagram}(c)) when the coupling is weak.
This result is due to the competition between interatomic interaction and
lattice coupling \cite{Hou20182}. The experimentally determined
spin-polarizations are in excellent, quantitative agreement with numerical
values (Fig.~\ref{fig:PhaseDiagram}(c-f)), indicating that the physically
realized system is consistent with the spin-momentum mixtures which result
in the superfluid stripe-phase observed in the numerics.

\begin{figure}[t]
\includegraphics[width=0.45\textwidth]{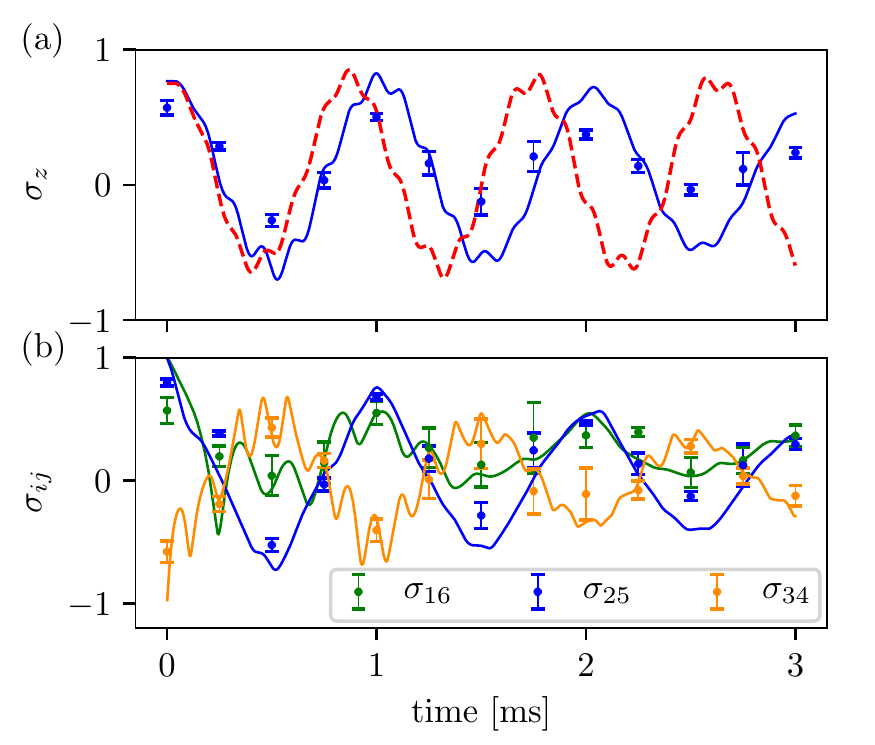}
\caption{(a) The Rabi oscillation of the total spin polarization after
suddenly jumping on the lattice. Experimental measurements are given by blue
points. Blue solid and red dashed lines represent numerical simulations with
and without interactions, respectively. (b) The Rabi oscillations between
individual spin-momentum channels corresponding to the bare-state basis as
marked in Fig.~\protect\ref{fig:experiment}(d). The data shown here
correspond to SOC strength $\hbar \Omega _{R}=2.7E_{R}$, detuning $\protect%
\delta =250$ Hz and lattice strength $\hbar \Omega _{L}=1.0E_{R}$.
Experimental data points are averages over four measurements with error bars
given by statistical error.}
\label{fig:RabiOscTot}
\end{figure}

{\color{blue}\emph{Coherent Rabi oscillation}}. The coherent nature of the
lattice-induced hopping between two band minima can be experimentally
demonstrated by observing Rabi oscillations induced by a sudden
quench of system parameters. After adiabatically dressing the BEC with SOC,
the lattice beams are jumped on, which initiates an oscillation between the two
SOC dispersion minima. After a certain evolution time, all beams are turned
off and the individual spin/momentum states are imaged. The resulting spin
polarization oscillates in time as shown in Fig.~\ref{fig:RabiOscTot}(a).
The observed oscillation frequency is in good agreement with the GPE simulation,
demonstrating the coherent dynamics of the system. Additionally, Fig.~\ref{fig:RabiOscTot}(b) shows the same
oscillation for particular spin-momentum state pairs as marked in Fig.~\ref%
{fig:experiment}(d), where $\sigma _{ij}=(N_{i}-N_{j})/(N_{i}+N_{j})$. We
see good agreement between the experimental values for each of the dominant
spin-momentum coupling channels and the corresponding GPE simulation. While
the vast majority of the atoms populate the bare states marked 2 and 5, Fig.~%
\ref{fig:RabiOscTot}(b) indicates that we are able to observe and resolve
the coherent oscillations in the coupled spin-momentum space beyond simple
SOC, in agreement with GPE simulation.

To explore the role of particle interactions, Fig.~\ref{fig:RabiOscTot}(a)
shows the results of GPE simulations with and without interactions. A
two-level single-particle Rabi oscillation has a frequency given by $\hbar
\omega =\sqrt{(\hbar \delta k_{L}/2k_{R})^{2}+\tilde{V}^{2}}$ where $\tilde{V%
}=\hbar \Omega _{L}\sqrt{1-(k_{L}/k_{R})^{2}}/2$ is the effective coupling
strength \cite{Supp}. With the experimental parameters described in Fig.~\ref%
{fig:RabiOscTot}, the single-particle Rabi oscillation period is evaluated
to be 0.8 ms, which is consistent with GPE simulation without interaction.
The role played by nonlinear effects from interatomic interaction is treated
on the mean-field level through a variational method \cite%
{Supp}. Intuitively, the density-density interaction, which dominates over
spin interaction in $^{87}$Rb, costs energy and thus makes it less favorable for the
condensate to simultaneously occupy both momentum states. As a result, we expect the
interactions reduce the frequency of oscillation and introduce damping
effects, which are confirmed by our GPE simulations. We observe strong
agreements between numerical simulations with interactions and experimental
observations, though the experimental data exhibit stronger damping which
is not captured by the mean-field analysis. Some higher frequency, lower
amplitude oscillations apparent in both single-particle and interacting
cases in Fig.~\ref{fig:RabiOscTot} are signatures of the highly detuned
couplings within the SOC dressed states. Those high-energy states are
coupled with different coupling strengths and are less populated, leading to
the small ripples on the overall Rabi oscillation. The Rabi oscillations
shown here are driven by a weak lattice which is slightly stronger than the
mean field interaction strength. This character may be indicative of the
dynamics underlying momentum-space Josephson junctions \cite{Hou20182},
which is beyond the scope of these experimental observations.

\begin{figure}[h]
\includegraphics[width=0.42\textwidth]{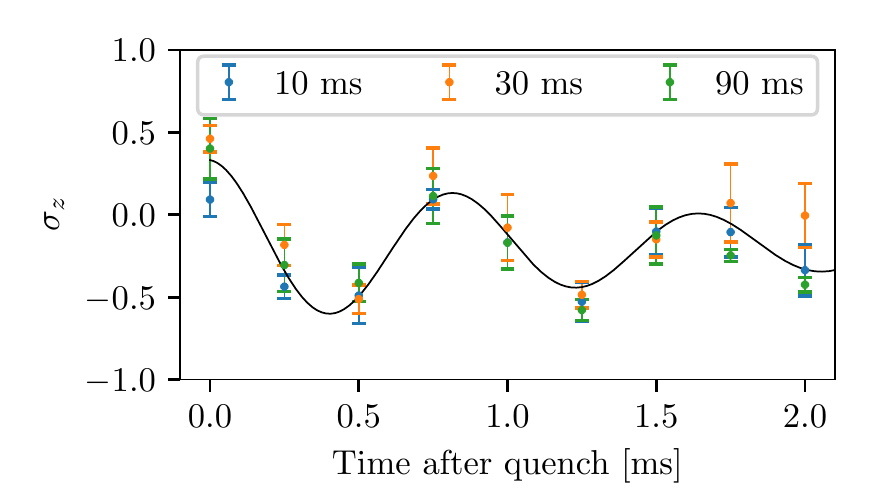}
\caption{Detuning quench induced oscillations. A lattice-coupled SOC
system is prepared in a ground state where $\hbar\Omega =2.7E_{R}$, $\delta =500$ Hz,
and $\Omega_L =1.0E_{R}$. The system is held in the ground state
 for 10 (blue), 30 (orange), or 90 ms (green) before the detuning is quenched from 500 Hz
  to -500 Hz. Experimental points are averages of four measurements with error bars
representing statistical errors and a least-squares fit to a damped harmonic
oscillation as a guide to the eye.}
\label{fig:DetQuench}
\end{figure}

{\color{blue}\emph{Stripe-phase ground state stability}}. While previous
arguments have indicated that we reliably prepare the appropriate
spin-momentum mixtures to achieve a high-contrast stripe-phase, those
mixtures must be coherent over a long period of time to produce the expected
density modulation for further investigation of dynamical properties and
excitations of superfluid stripe phases. To study the coherent lifetime of
the ground stripe state, we employ quenches of system parameters initiated
at various wait times after the preparation of the ground state. The
reproducibility of the ground-state quench dynamics provides evidence for
the long time phase stability of the spin-momentum mixtures which produces
the stripe-phase.

An example of such a quench is a sudden jump of the SOC detuning $\delta $,
which induces coherent oscillations of the spin polarization and
corresponding momentum states. We begin by preparing the superfluid
stripe-phase with $\approx 2\times 10^{5}$ atoms and quench $\delta $ from
500 Hz to $-$500 Hz. We perform the quench 10, 30, or 90 ms after the preparation
of the ground state and in all cases observe
subsequent oscillations of similar amplitude and frequency. Fig.~\ref%
{fig:DetQuench} shows the spin-polarization time series after each of the
wait times along with a best-fit curve as a guide to the eye. This
consistency indicates that the ground state supported by the lattice-coupled
SOC is stable in phase and population on the order of at least 100 ms,
providing the possibility for investigating long-time dynamics of
supersolid-like phases.

{\color{blue}\emph{Conclusions}}. In this work, we have demonstrated a
robust framework for the production of a high-contrast, long-lived, and
tunable ground state stripe-phase in a SOC BEC. By showing quantitative
agreement with numerical simulations of the GPE, we have demonstrated that
our experimental configuration is able to reliably produce the expected
stripe-phase over a broad parameter space. We confirm that the experimental
momentum and spin distributions agree with those expected for both the
static and dynamic cases and are coherent over long times ($\sim $100 ms).

The application of optical lattice driven momentum-space hopping allows for
the SOC stripe-phase to exist over a broad parameter space comparing to
previous experiments. Therefore, this configuration opens up new
possibilities for investigations of excitation dynamics such as roton modes
and breathing modes as well as studies of defects or barriers passing
through the superfluid stripe-phase. Additionally, we note that such a setup
can also be used to realize momentum-space Josephson effects when the
lattice coupling strength is weak compared to particle interactions \cite%
{Hou20182}.

{\color{blue}\emph{Acknowledgements}}. TMB, SM, VG and PE acknowledge
funding from the NSF (PHY-1607495). JH, XL, KS, and CZ are thankful for
support from NSF (PHY-1505496), AFOSR (FA9550-16-1-0387) and ARO
(W911NF-17-1-0128).


\newpage \clearpage
\onecolumngrid
\appendix

\section{Supplementary Materials}

\section{Momentum space hopping induced by optical lattices}

The analysis starts with a zero-detuned, $\delta =0$, spin-orbit coupled
system with the usual quasimomentum rotation \cite{Lin2011} $q=k\pm k_{R}$.
Throughout this Supplementary Material, we use dimensionless units for
convenience. We apply a lattice potential,
\begin{equation}
V_{\mathrm{lat}}(x)=2\Omega _{L}\sin
^{2}k_{L}x=-V(e^{2ik_{L}x}+e^{-2ik_{L}x})+\mathrm{{const}.,}
\end{equation}%
where $V=\Omega _{L}/2$. The potential induces the first-order coupling
between states with quasimomenta $q$ and $q\pm 2k_{L}$. The complete $6$%
-level Hamiltonian is given by
\begin{equation}
H=\left( {%
\begin{array}{cccccc}
{{{(q-{k_{R}}+2{k_{L}})}^{2}}} & V & 0 & 0 & 0 & 0 \\
V & {{{(q-{k_{R}})}^{2}}} & V & {\frac{\Omega _{R}}{2}} & 0 & 0 \\
0 & V & {{{(q-{k_{R}}-2{k_{L}})}^{2}}} & 0 & {\frac{\Omega _{R}}{2}} & 0 \\
0 & {\frac{\Omega _{R}}{2}} & 0 & {{{(q+{k_{R}})}^{2}}} & V & 0 \\
0 & 0 & {\frac{\Omega _{R}}{2}} & V & {{{(q+{k_{R}}-2{k_{L}})}^{2}}} & V \\
0 & 0 & 0 & 0 & V & {{{(q+{k_{R}}-4{k_{L}})}^{2}}}%
\end{array}%
}\right) ,
\end{equation}%
where the first three basis states are spin-up and the latter ones are
spin-down, in sequential order. In Fig.~\ref{fig:experiment}(c) we label the
6 states of the Hamiltonian basis on the real-spin branches and show
couplings between them.

The ground state band of the SOC system possesses two minima at
\begin{equation}
q_{\min}=\pm \sqrt{1-\left(\frac{\Omega _{R}}{4}\right)^{2}}.
\end{equation}%
Consequently, we require $k_{L}=q_{\min }$ so that the lattice coupling
connects the SOC band minima. By using the above relation defining
$k_{R}=1$ as the unit of momentum, the Hamiltonian becomes
\begin{equation}
H=\left( {%
\begin{array}{cccccc}
{{{(3{k_{L}}-1)}^{2}}} & V & 0 & 0 & 0 & 0 \\
V & {{{({k_{L}}-1)}^{2}}} & V & {2\sqrt{1-k_{L}^{2}}} & 0 & 0 \\
0 & V & {{{({k_{L}}+1)}^{2}}} & 0 & {2\sqrt{1-k_{L}^{2}}} & 0 \\
0 & {2\sqrt{1-k_{L}^{2}}} & 0 & {{{({k_{L}}+1)}^{2}}} & V & 0 \\
0 & 0 & {2\sqrt{1-k_{L}^{2}}} & V & {{{({k_{L}}-1)}^{2}}} & V \\
0 & 0 & 0 & 0 & V & {{{(3{k_{L}}-1)}^{2}}}%
\end{array}%
}\right) .
\end{equation}

We can rewrite the Hamiltonian in a pseudo-spin basis that diagonalizes the
spin-orbit Hamiltonian through a unitary transformation
\begin{equation}
U^{\dagger }HU=H_{d}+VH_{nd},
\end{equation}%
with $H_{d}$ diagonalized in the transformed basis such that
\begin{equation}
H_{d}=\text{diag}\left(
(3k_{L}-1)^{2},~k_{L}^{2}-1,~k_{L}^{2}+3,k_{L}^{2}+3,~k_{L}^{2}-1,~(3k_{L}-1)^{2}\right) ,
\end{equation}%
and the tunneling terms appear in the off-diagonal part of the Hamiltonian,
\begin{equation}
H_{nd}=\left(
\begin{array}{cccccc}
0 & -\frac{\sqrt{k_{L}+1}}{\sqrt{2}} & 0 & \frac{\sqrt{1-k_{L}}}{\sqrt{2}} &
0 & 0 \\
-\frac{\sqrt{k_{L}+1}}{\sqrt{2}} & 0 & -k_{L} & 0 & \sqrt{1-k_{L}^{2}} & 0
\\
0 & -k_{L} & 0 & \sqrt{1-k_{L}^{2}} & 0 & \frac{\sqrt{1-k_{L}}}{\sqrt{2}} \\
\frac{\sqrt{1-k_{L}}}{\sqrt{2}} & 0 & \sqrt{1-k_{L}^{2}} & 0 & k_{L} & 0 \\
0 & \sqrt{1-k_{L}^{2}} & 0 & k_{L} & 0 & \frac{\sqrt{1+k_{L}}}{\sqrt{2}} \\
0 & 0 & \frac{\sqrt{1-k_{L}}}{\sqrt{2}} & 0 & \frac{\sqrt{1+k_{L}}}{\sqrt{2}}
& 0 \\
&  &  &  &  &
\end{array}%
\right) .
\end{equation}

\begin{figure}[h]
\includegraphics[width=2.25in]{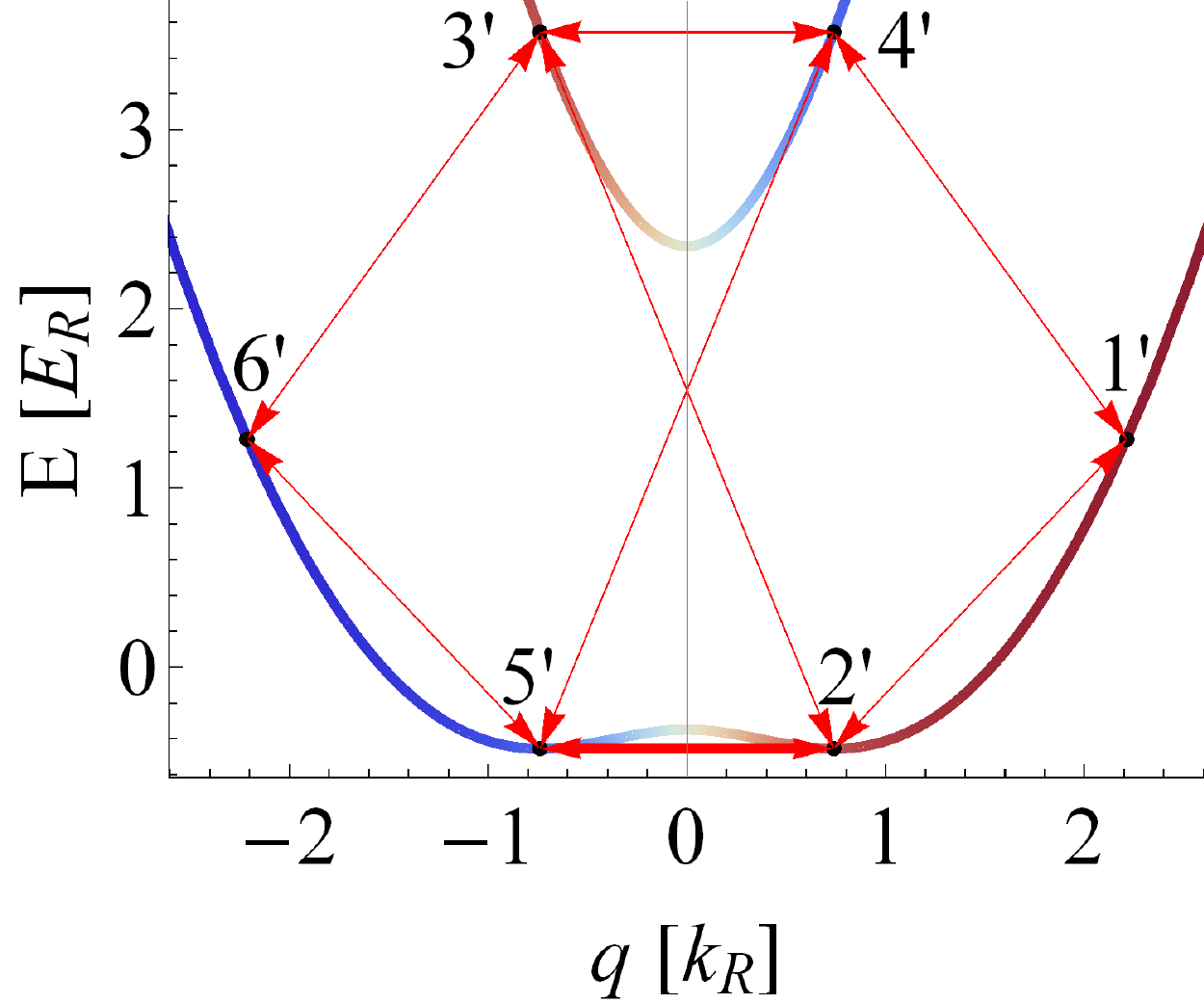}
\caption{Dressed lattice coupling diagram. The color coding indicates the
relative spin populations where red is predominantly spin-up and blue is
strongly spin-down.}
\label{fig:dressedcoupling}
\end{figure}

The couplings between different states in this pseudo-spin basis are
illustrated in Fig.~\ref{fig:dressedcoupling}. In this basis, $|2^{\prime
}\rangle $ is coupled directly to $|5^{\prime }\rangle $ with a strength $%
\sqrt{1-k_{L}^{2}}V$ (the two upper band states, $|3'\rangle$ and $|4'\rangle$, are coupled with the same
strength), while $|2'\rangle$ couples to higher excited states with stengths $-\sqrt{\frac{%
1+k_{L}}{2}}V$ to $|1'\rangle$ and $-k_{L}V$ to $|3'\rangle$. Consequently, we expect larger coupling
between $|2^{\prime }\rangle $ and $|5^{\prime }\rangle $ when $k_{L}$ is
small. If detuning is added into the analysis, we find it scales as $%
k_{L}\delta $ in the eigenenergies leads to a coupling of states $|2^{\prime }\rangle $
to $|4^{\prime }\rangle $ and $|3^{\prime }\rangle $ to $|5^{\prime }\rangle
$, i.e., from the lower band to the upper band directly, with a strength $\frac{1}{2}%
\sqrt{1-k_{L}^{2}}\delta $. This term is dropped since $\delta \ll \Omega
_{L}$ in the dynamics of interest.

\section{Density modulation in ground state}

\begin{figure}[h]
\includegraphics{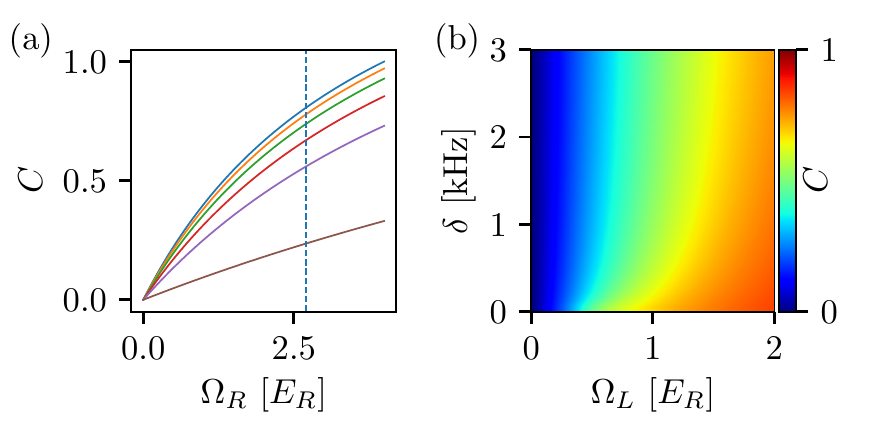}
\caption{(a) Fringe contrast for a variety of mixed momentum populations
from $n_{1}/n_{2}=\{1,2,3,5,10,100\}$, from top to bottom, as a function of
SOC strength. A dotted line marks the coupling strength used in this work, $%
\Omega _{R}=2.7E_R$. (b) The fringe contrast as a function of SOC detuning $%
\protect\delta $ and lattice coupling strength $\Omega _{L}$, where the SOC
strength is fixed at $\Omega _{R}=2.7E_R$. }
\label{fig:NumGroundState}
\end{figure}

A rough analytic model for the SOC stripe-phase contrast can be derived from
\cite{li2012quantum} where the number density as a function of position is a
vertically offset cosine curve with amplitude $n_{A}=2\sqrt{n_{1}n_{2}}%
\Omega _{R}/(4E_{R})$. Then the contrast is defined as $C=2n_{A}/(1+n_{A})$,
where $n_{1},n_{2}$ are the fractional density of two momentum states and $%
n_{1}+n_{2}=1$. In Fig.~\ref{fig:NumGroundState}(b) we plot the contrast as
a function of the Raman detuning for a wide range of $n_{1}/n_{2}$. At any
given Raman coupling, the contrast is maximized when $n_{1}=n_{2}$. At high
Raman coupling, a sizable modulation occurs even when the minima are
unevenly populated. This is also consistent with our numerical simulations
shown in Fig.~\ref{fig:NumGroundState}(c), which give the phase diagram of the
contrast $C$.

\section{Two-state model and quench dynamics}

In this section, we study population and phase oscillations between $%
|2^{\prime }\rangle $ and $|5^{\prime }\rangle $ as marked in Fig.~\ref%
{fig:dressedcoupling}. For simplicity, we ignore higher-energy states and
consider an effective two-level system. To simplify the notation we also
drop the primes in the following discussion. The validity of this 2-level
model has been studied thoroughly in previous work \cite{Hou20182}. The
system dynamics can be approximately described by%
\begin{equation}
i\partial _{t}\left( {%
\begin{array}{c}
C_{2} \\
C_{5}%
\end{array}%
}\right) =(H_{0}^{\text{eff}}+H_{I}^{\text{eff}})\left( {%
\begin{array}{c}
C_{2} \\
C_{5}%
\end{array}%
}\right) ,
\end{equation}%
where the effective single particle Hamiltonian $H_{0}^{\text{eff}}$ is
given by
\begin{equation}
H_{0}^{\text{eff}}=-\tilde{\delta}{\tau _{z}}+\tilde{V}{\tau _{x}}=\omega
(-\cos \alpha ~{\tau _{z}}+\sin \alpha ~{\tau _{x}})
\end{equation}%
with $\tilde{\delta}=\delta k_{L}/2$, $\tilde{V}=V\sqrt{1-k_{L}^{2}}$, $\tan
\alpha =\tilde{V}/\tilde{\delta}$ and $\{\tau \}$ are Pauli matrices. The
interaction part is
\begin{equation}
H_{I}^{\text{eff}}=2g_{G}\left( {%
\begin{array}{cc}
|C_{5}|^{2} & 0 \\
0 & |C_{2}|^{2}%
\end{array}%
}\right) ,
\end{equation}%
where $g_{G}=ng(1-k_{L}^{2})$. Rewriting $C_{j}=N_{j}e^{i\theta _{j}},j=2,5$%
, we can recast the 2-level model into two classical equations of motion
\begin{eqnarray}
\partial _{t}z &=&-\sqrt{1-z^{2}}\sin \phi , \\
\partial _{t}\phi &=&\frac{g_{G}}{\tilde{V}}z+\frac{z}{\sqrt{1-z^{2}}}\cos
\phi +\frac{1}{2}\frac{k_{L}\delta }{\tilde{V}},
\end{eqnarray}%
where $z=(N_{2}-N_{5})/(N_{2}+N_{5})$, $\phi =\theta _{2}-\theta _{5}$ and
time has been rescaled according to $t\rightarrow 2\tilde{V}t$. To observe
interaction effects like self-trapping, we typically require $\tilde{\delta}/%
\tilde{V}\ll 1$ and $g_{G}/\tilde{V}\gg 1$, which are outside the parameter
space we consider in the experiments. When $g_{G}/\tilde{V}\ll 1$, we simply
have a coherent Rabi oscillation with frequency $\omega =\sqrt{\tilde{\delta}%
^{2}+\tilde{V}^{2}}$. In the experiment, we study the dynamics mainly in the
intermediate region (i.e., $\Omega _{L}\gtrsim 1$ and consequently $g_{G}/%
\tilde{V}\lesssim 0.5$). Although strong interaction effects cannot be
observed, changes in the periodicity still reveal the important role played
by interaction in the dynamics.

\end{document}